\documentclass[twocolumn,prl,aps,showpacs,footinbib,amsmath,amssymb,superscriptaddress,longbibliography]{revtex4-2}
\usepackage{graphicx}
\usepackage{dcolumn}
\usepackage{xcolor}
\usepackage{bm}
\usepackage{hyperref}
\usepackage{amsmath,amssymb,amsfonts,bbold}
\usepackage[normalem]{ulem}
\usepackage{miller}
\usepackage{todonotes}
\usepackage{physics}
\usepackage[thinspace,amssymb]{SIunits}
\usepackage{booktabs}
\usepackage{tabstackengine}
\TABstackMath
\graphicspath{{figures/}}

\usepackage{natbib}

\definecolor{dartmouthgreen}{rgb}{0.05, 0.5, 0.06}
\definecolor{darkspringgreen}{rgb}{0.09, 0.45, 0.27}
\definecolor{DebianRed}{rgb}{0.84, 0.04, 0.33}
\definecolor{darkpowderblue}{rgb}{0.0, 0.2, 0.6}
\definecolor{lightgreen}{rgb}{0.0, 1.0, 0.0}

\definecolor{orange}{HTML}{FF8000}
\definecolor{lightgray}{HTML}{C0C0C0}
\definecolor{nightblue}{HTML}{001080}
\definecolor{darkred}{HTML}{CB0505}

\newcommand{\didu}{$\dd I/\dd U$}
\newcommand{\ddidu}{$\dd^2I/\dd U^2$}

\begin{document}

\title{Spin-resolved spectroscopic evidence for spinarons in Co adatoms}

\author{Felix Friedrich}
\author{Artem Odobesko}
\email[]{Corresponding author for the experiments: artem.odobesko@physik.uni-wuerzburg.de}
\affiliation{Physikalisches Institut, Experimentelle Physik II, 
	Julius-Maximilians-Universit\"at W\"urzburg and W\"urzburg-Dresden Cluster of Excellence ct.qmat, 
	Am Hubland, 97074 W\"urzburg, Germany}
\author{Juba Bouaziz}
\affiliation{Peter Gr\"unberg Institut and Institute for Advanced Simulation, 
	Forschungszentrum J\"ulich and JARA, 52425 J\"ulich, Germany}
\author{Samir Lounis}
\email[]{Corresponding author for the theory: s.lounis@fz-juelich.de}
\affiliation{Peter Gr\"unberg Institut and Institute for Advanced Simulation, 
	Forschungszentrum J\"ulich and JARA, 52425 J\"ulich, Germany}
\affiliation{Faculty of Physics, University of Duisburg-Essen and CENIDE, 47053 Duisburg, Germany}
\author{Matthias Bode}
\affiliation{Physikalisches Institut, Experimentelle Physik II, 
	Julius-Maximilians-Universit\"at W\"urzburg and W\"urzburg-Dresden Cluster of Excellence ct.qmat, 
	Am Hubland, 97074 W\"urzburg, Germany}
\affiliation{Wilhelm Conrad R\"ontgen-Center for Complex Material Systems (RCCM), 
	Julius-Maximilians-Universit\"at W\"urzburg, Am Hubland, 97074 W\"urzburg, Germany}

\date{\today}

\begin{abstract}
Single cobalt atoms on the (111) surfaces of noble metals were for a long time considered prototypical 
systems for the Kondo effect in scanning tunneling microscopy experiments. 
Yet, recent first-principle calculations suggest that the experimentally observed spectroscopic zero-bias anomaly (ZBA)
should be interpreted in terms of excitations of the Co atom's spin and the formation of a novel quasiparticle, the spinaron, 
a magnetic polaron resulting from the interaction of spin excitations with conduction electrons, rather than in terms of a Kondo resonance.
Here we present state-of-the-art spin-averaged and spin-polarized
scanning tunneling spectroscopy measurements on Co atoms on the Cu(111) surface in magnetic fields of up
to 12 T, that allow us to discriminate between the different theoretical models 
and to invalidate the prevailing Kondo-based interpretation of the ZBA. 
Employing extended \textit{ab-initio} calculations, we instead provide strong evidence for multiple spinaronic states in the system.
Our work opens a new avenue of research to explore the characteristics and consequences 
of these intriguing hybrid many-body states as well as their design in man-made nanostructures.
\end{abstract}


\maketitle

The collective screening of an impurity spin by conduction electrons in the Kondo effect 
is one of the most fundamental manifestations of many-body phenomena in physics~\cite{kondo1964}. 
Resulting in a resistance minimum at the Kondo temperature in transport measurements of magnetically doped samples, 
the experimental identification of the Kondo effect in individual atoms is linked to the rise of a local electronic state 
at the Fermi level with a characteristic temperature and field dependence \cite{abrikosov1965,suhl1965,ternes2009}. 
Corresponding spectroscopic signatures 
with a zero-bias anomaly (ZBA) in the differential conductance signal
were for the first time observed in scanning tunneling spectroscopy (STS) experiments on Ce~\cite{li1998} 
and Co atoms~\cite{madhavan1998,manoharan2000,knorr2002,schneider2002} on noble metal (111) surfaces. 
While the ZBA detected on Ce/Ag(111) was later shown to result from vibrational excitations 
of a hydrogen atom adsorbed to the Ce atom~\cite{pivetta2007}, 
the anomaly found atop Co atoms on noble metal (111) surfaces is still explained by a Fano resonance 
caused by interfering tunneling paths into the Kondo state and atomic orbitals~\cite{fano1961,fernandez2021,tacca2021}. 

However, a different interpretation of the ZBA on Co atoms on the noble metal (111) surfaces was recently  proposed
in terms of spin excitations (SEs) and their interaction with conduction electrons, 
leading to the formation of a localized quasiparticle, 
a magnetic polaron that was termed ``spinaron''~\cite{schweflinghaus2014,bouaziz2020}. 
This intriguing feature can be interpreted as the atomically localized version of magnetic polarons 
emerging from the coupling of magnons and electrons or holes~\cite{edwards1973}.

In this contribution we investigate Co atoms on the Cu(111) surface in magnetic fields of up to \unit{12}{T} 
by means of spin-averaged and spin-polarized STS, to experimentally assess this controversy. 
We find that the ZBA on Co/Cu(111) consists of multiple spectroscopic features that show a distinct field-dependent energy shift. 
Spin-resolved data reveal that features with majority (minority) spin character 
shift to higher (lower) energies with increasing field strength, 
a behavior which is opposite to the conventionally expected field-dependent behavior.
These findings invalidate the Kondo effect as the origin of the ZBA, 
but instead provide evidence for the presence of the predicted spinaron.
This interpretation is supported by  first-principles calculations 
based on time-dependent density functional theory (TD-DFT) and many-body perturbation theory (MBPT). 
Our findings are a crucial milestone in all nano-physics fields gravitating around the interplay 
of many-body effects, strong correlations and magnetic properties. 

\vspace{-0.3cm}
\subsection*{Field dependence of magnetic states and excitations}
\vspace{-0.4cm}
In the past, significant experimental and theoretical effort was made to investigate 
the properties of the ZBA in Co atoms on various surfaces. 
For example, the influence of the Shockley-like surface state present on noble metal (111) 
surfaces~\cite{barral2004,lin2006,henzl2007,baruselli2015,li2018,moro-lagares2018,fernandez2021,tacca2021}, 
or of exchange~\cite{madhavan2001,neel2008,kawahara2010,choi2016a,neel2020} 
as well as spin-orbit coupling~\cite{shick2022} onto the presumed Kondo resonance were studied systematically. 
In contrast, two properties characteristic for the Kondo effect 
have not yet been demonstrated for Co atoms on the noble metal (111) surfaces: 
First, the broadening of the observed resonance at temperatures 
above the Kondo temperature $T_\mathrm{K}$ \cite{hanke2005,otte2008} could not be shown, 
because Co atoms on Cu, Ag and Au(111) become highly mobile 
at temperatures much lower than $T_\mathrm{K}$~\cite{knorr2002b}.  
Second, no magnetic field dependence of the tunneling conductance signal 
has been experimentally reported to date~\cite{patton2007,seridonio2009}, 
contrary to, e.g., Co atoms on $\mathrm{Cu}_2\mathrm{N}$ layers~\cite{otte2008,vonbergmann2015}, 
where a clear splitting of the ZBA was observed. 
Yet, as will be shown below, it is precisely this field dependence which allows us to unambiguously discriminate 
between the various explanations for the ZBA and ultimately provides evidence for spinarons in the system.

\begin{figure}[t]
\includegraphics[width=88mm]{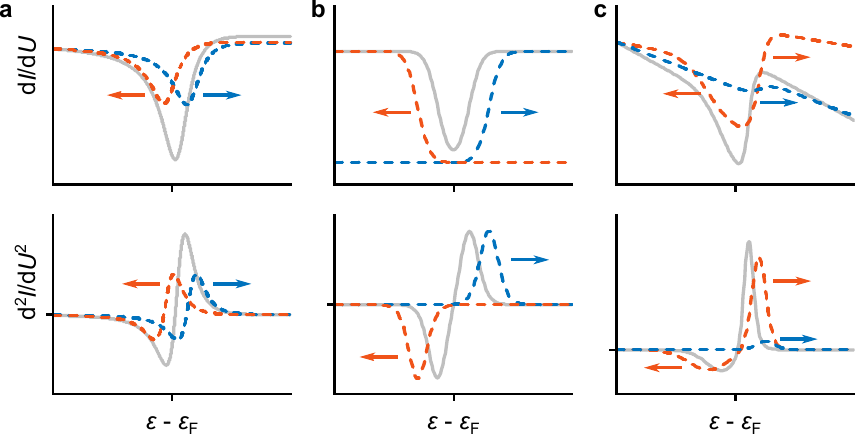}
\caption{\textbf{Schematic representation of the field-induced splitting of magnetic resonances and excitations.} 
	Gray lines show the zero-field \didu\ (upper panels) and \ddidu\ spectra (lower panels), red (blue) dashed lines represent 
	the majority (minority) contributions of the magnetic field-split resonances/excitations. 
	Arrows indicate the direction of the shift with increasing field strength. 
	\textbf{a} Fano curve 
	originating from a Kondo resonance. 
	\textbf{b} Bias-symmetric steps which reflect the inelastic tunneling of electrons inducing an excitation of the system's spin. 
	Equal to the features in a Kondo system, majority (minority) SEs shift to lower (higher) energies. 
	\textbf{c} Contrary, a spinaron, which prominently appears as a peak in the \ddidu\ signal of the majority channel, 
	shifts to higher energies despite having majority spin character, while all other features representing SEs, shift as presented in b.}
\label{fig:fig1}
\end{figure}

Figure \ref{fig:fig1} schematically illustrates the effect of an external magnetic field on tunneling spectra 
expected for the three potential effects that might cause the ZBA in Co/Cu(111): 
(i) a Kondo resonance, (ii) inelastic SEs, and (iii) the newly proposed model of SEs in combination with the spinaron.
(i) As the Kondo resonance splits in an external magnetic field, the observed Fano curves  
equally split symmetrically away from their zero-field position, see Fig.~\ref{fig:fig1}a.
Majority (red) and minority (blue) components shift to lower and higher energies, respectively~\cite{vonbergmann2015}. 
As can be seen from Fig.~\ref{fig:fig1}b, (ii) SEs caused by the inelastic tunneling of electrons 
lead to steps in the differential conductance which show a similar magnetic field-dependent behavior. 
Like the Fano curves of a Kondo system, the steps in the majority (minority) spin channel 
shift to lower (higher) energies~\cite{loth2010a,loth2010}. 
This behavior was also confirmed for more intriguing manifestations of SEs found in recent tunneling spectroscopy experiments~\cite{brinker2022}.

The scheme in Fig.~\ref{fig:fig1}c displays what is expected for (iii) the combination of SEs and the spinaron~\cite{bouaziz2020}.
Within a simple scenario, the response of the spinaron to the magnetic field 
is expected to be opposite to that of SEs in the same spin channel. 
The spinaron is a magnetic polaronic state, involving the hybrid of a SE and an electron or hole 
with a spin character \textit{opposite} to that of the spin channel hosting the spinaron. 
This electron (hole) induces a potential which is capable of binding the hybrid state forming the spinaron. 
As for the other conventional magnetic states, the majority (minority) spin potential 
shifts to lower (higher) energies in the external magnetic field, 
but consequently drags the bound hybrid spinaronic state in the opposite energetic direction. 
Therefore, the spinaron found in the majority spin channel in the calculations of Ref.~\cite{bouaziz2020} 
is expected to move to higher energies with increasing field strength, 
i.e.\ in the opposite direction than any other conventional feature in the same spin channel. 
The additional intrinsic SEs predicted for Co/Cu(111) will follow the trend presented in (ii).
This fundamental difference is exploited in the following experiments to determine the physical mechanism 
which causes the ZBA in Co atoms on the noble metal (111) surfaces.

\subsection{Splitting of the zero-bias anomaly in an external magnetic field}
\vspace{-0.3cm}

\begin{figure*}
\includegraphics[width=180mm]{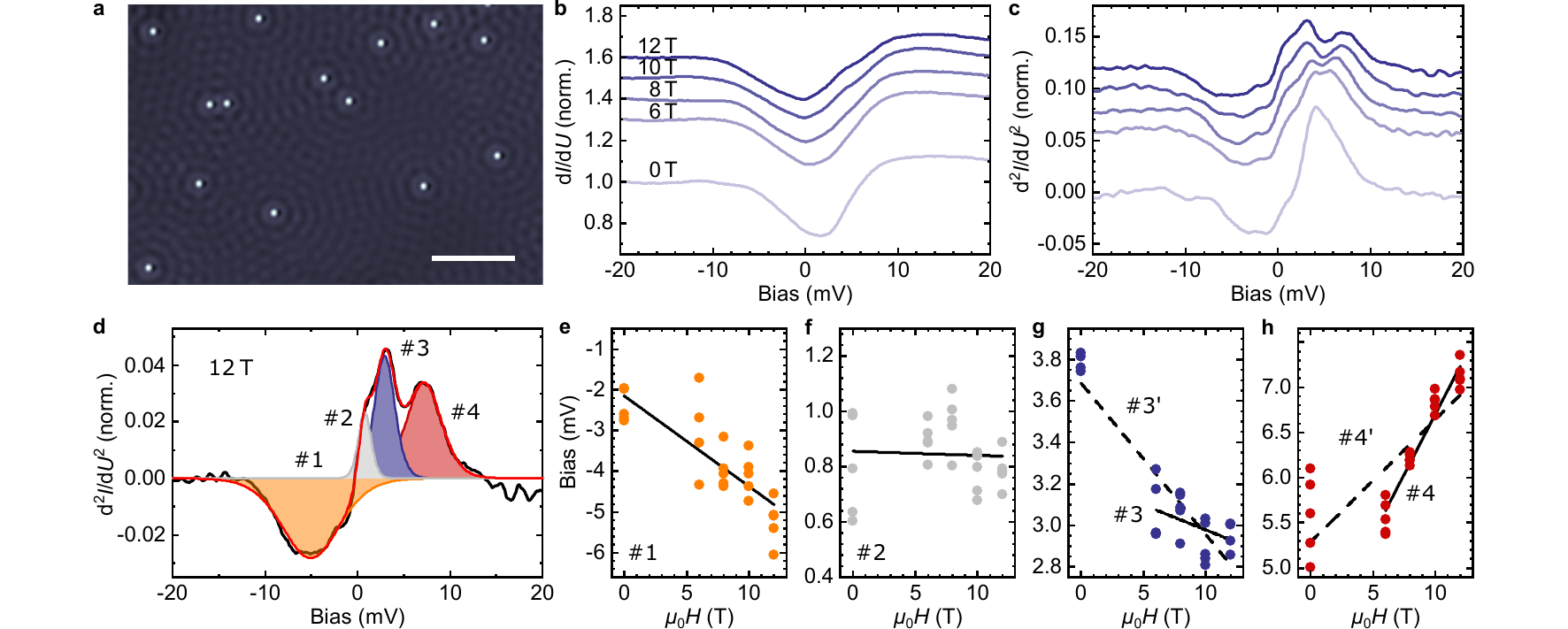}
\caption{\textbf{Magnetic field-dependent splitting of the zero-bias anomaly.} 
	\textbf{a} Scan of the Cu(111) surface with single adsorbed Co atoms. Scale bar is \unit{10}{nm}. 
	\textbf{b} Differential conductance \didu\ spectra of a single Co atom at indicated fields. 
	Spectra are normalized and offset by \unit{0.05}{\mu_0H/T}. 
	\textbf{c} Numerically calculated derivative of the data presented in b. Curves are offset by \unit{0.01}{\mu_0H/T}. 
	\textbf{d} Exemplary fit of the \ddidu\ signal at \unit{12}{T} consisting of four Gaussians \#1--\#4 as described in the main text (red line). 
	\textbf{e-h} Energy of \#1--\#4 as a function of the applied field as extracted from measurements on five individual atoms. 
	Lines indicate linear fits to the data, where for peaks $\#3$ and $\#4$ fits including ($\#3'/\#4'$) and excluding ($\#3/\#4$) the zero-field data are displayed.
	Stabilization parameters: \textbf{a} \mbox{$U = \unit{-20}{mV}$,} \mbox{$I = \unit{0.4}{nA}$}; 
	\textbf{b-d} \unit{-20}{mV}, \unit{1}{nA}.} \label{fig:fig2}
\end{figure*}

To investigate the magnetic field-dependent splitting of the ZBA, 
we deposit Co atoms onto clean Cu(111) (see Methods). The resulting surface is displayed in Fig.~\ref{fig:fig2}a, 
where individual Co atoms surrounded by 
Friedel oscillations can be identified. 
Spectra recorded atop single Co atoms in external out-of-plane magnetic fields of up to \unit{12}{T} 
are presented in Fig.~\ref{fig:fig2}b for one representative Co atom~\cite{supplmat}. 
At zero magnetic field, we observe the same Fano resonance-like spectrum already reported in Ref.~\cite{manoharan2000}. 
With increasing magnetic field strength, the minimum of the differential conductance signal 
appears to shift to lower bias voltages and the overall width of the ZBA increases. 
At fields $\geq \unit{10}{T}$, a kink in the \didu\ signal at about \unit{+3}{mV} becomes visible. 

To emphasize these changes of the ZBA, we present the derivative of the differential conductance, the \ddidu\ signal, in Fig.~\ref{fig:fig2}c. 
The falling edge of the ZBA at negative bias voltages is translated into a dip, and the rising edge appears as a peak. 
The field-induced kink in the \didu\ signal at positive bias is reflected in the splitting of the \ddidu\ peak. 
To track the magnetic field-induced shift of the individual features, we fit the \ddidu\ signal by four Gaussian functions (\#1--\#4),
as exemplarily presented for the \unit{12}{T} data in Fig.~\ref{fig:fig2}d~\cite{supplmat}. Please note that the choice of a Gaussian function is not motivated by the underlying physics, but merely the shape of the observed features. 
The first Gaussian (\#1, orange) describes the dip at negative bias and the envelope of Gaussians \#3/\#4 (blue/red) 
accounts for the peak at positive bias voltage which splits at high fields. 
To fully capture the data, it is necessary to add the fourth Gaussian (\#2, gray) 
to account for the shoulder close to zero bias, that appears in all spectra. 
This fit enables a reasonable description of the experimental data, 
sufficient to extract the field-dependent position of the features. 
The respective energy positions of features \#1--\#4 displayed in Fig.~\ref{fig:fig2}e-h, respectively, 
can be fitted with a linear dependence on the magnetic field. 
The extracted values of the slope $\dd U/\dd(\mu_0H)$ 
and the intersection with the $y$-axis $U(H=0)$ are pre\-sen\-ted in Tab.~\ref{tab1}. 

In contrast to what is expected for a Kondo resonance or SEs, 
where all features are shifted by the same absolute value to higher or lower energies depending on their spin character,
the features detected on Co/Cu(111) exhibit a different behavior. 
We find that the \ddidu\ dip at negative bias (orange) and the outer peak at positive bias (red) 
exhibit a very similar absolute shift of $\approx$\,\unit{0.25}{mV/T}, 
but in opposite directions (cf.\ Fig~\ref{fig:fig2}e,h and Tab.~\ref{tab1}).
The peak at lower positive bias (blue), however, shifts with a much weaker field dependence of \unit{-0.07}{mV/T}  to lower energies (\unit{-0.02}{mV/T} if the zero-field data is excluded from the fit). 
Although different field dependencies are also observed for SEs 
whenever spin states are mixed due to a transverse anisotropy~\cite{hirjibehedin2007}, 
in this case all features should appear symmetrically around zero bias, which we do not observe in our experiments. 
Lastly, the peak close to zero bias does not show any field dependence, 
leading to the conclusion that it is of non-magnetic origin and can therefore be neglected in the following discussion. While no further investigations were performed to assess the origin of this feature, we speculate about a phononic mode of the adatom--surface system~\cite{cui1993}, acknowledging that the potential related negative-bias inelastic tunneling feature could be hidden by peak~\#1 of the ZBA.

\begin{table} 
	\caption{Magnetic field induced shift and zero-field energy of the four separate \ddidu\ features 
  		of the ZBA extracted from the linear fits displayed in Fig.~\ref{fig:fig2}e-h.} \label{tab1} 
	\begin{center}
		\begin{tabular}{|c|c|c|} \hline Peak \# & Shift (mV/T) & Position at \unit{0}{T} (mV) \\ 
		\hline 1~$\textcolor{orange}{\bullet}$ 	& $-0.22 \pm 0.03$ 		& $-2.2 \pm 0.3$ 	\\ 
			2~$\textcolor{lightgray}{\bullet}$	& $-0.001 \pm 0.006$	& $0.86 \pm 0.05$    \\ 
			3~$\textcolor{nightblue}{\bullet}$	    & $-0.024 \pm 0.011$	& $3.2 \pm 0.1$    \\
			3'~$\textcolor{nightblue}{\bullet}$	& $-0.073 \pm 0.007$	& $3.69 \pm 0.06$    \\
			4~$\textcolor{darkred}{\bullet}$		& $0.27 \pm 0.02$		& $4.0 \pm 0.1$    \\
			4'~$\textcolor{darkred}{\bullet}$		& $0.14 \pm 0.01$		& $5.3 \pm 0.2$    \\
		\hline 
		\end{tabular}
	\end{center} 
\end{table}

The field-dependent data presented in Fig.~\ref{fig:fig2} is clearly inconsistent with the simple Kondo picture, 
as we find features exhibiting different energy shifts as a function of the field strength. 
We can also exclude that the ZBA is caused by pure SEs, as the observed features are not symmetric with respect to zero bias. 
Yet, the origin of the ZBA still remains elusive. Especially a direct hint at the predicted spinaron is so far missing.

\subsection{Spin polarization of the zero-bias anomaly}
\vspace{-0.3cm}

\begin{figure}[t]
\includegraphics[width=88mm]{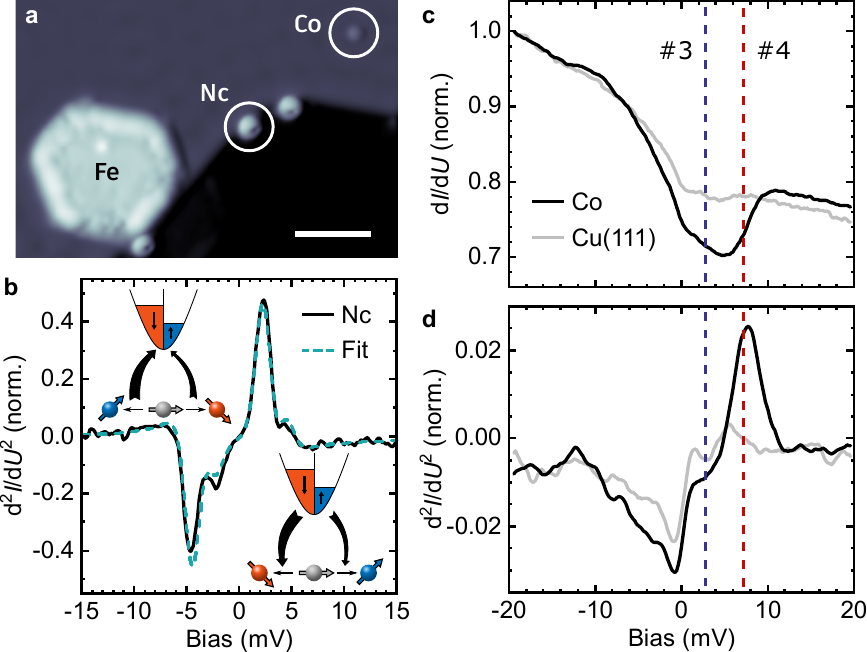}
\caption{\textbf{Implementation and results of spin-resolved measurements on Co/Cu(111). 
	a} Topographic scan of the Cu(111) surface decorated with Fe islands, Nc molecules and Co atoms. 
	After picking up a small magnetic Fe cluster, the tip polarization is determined 
	by measuring the SE signal on top of a Nc molecule. 
	\textbf{b} A fit to the data reveals a spin polarization of the tip of $\approx$ \unit{60}{\%}. 
	Insets illustrate the spin excitation process detected with a spin-polarized tip. 
	\textbf{c} Differential conductance \didu\ and \textbf{d} the \ddidu\ signal 
	measured atop a Co atom (black) and the bare Cu(111) surface (gray) with the same tip. 
	Dashed lines indicate the position of the blue and red \ddidu\ feature as determined in Fig.~\ref{fig:fig2}.
	Stabilization parameters: $\mu_0H = \unit{12}{T}$; \textbf{a},\textbf{b} \mbox{$U = \unit{-20}{mV}$,} \mbox{$I = \unit{0.1}{nA}$}; 
	\textbf{c},\textbf{d} \unit{-20}{mV}, \unit{0.4}{nA}. Scale bar is \unit{4}{nm}.} \label{fig:fig3}
\end{figure}

To elucidate the physical origin of the ZBA, we performed spin-polarized STS measurements 
to identify the spin character of the observed spectroscopic features. 
For this purpose, small magnetic clusters are transferred 
to the tip apex by dipping the tip into Fe islands grown on the clean Cu(111) surface. 
The tip magnetization is subsequently determined by assessing SEs 
in nickelocene (Nc) molecules on the same surface (see Methods for Fe and Nc deposition). 
Nc was shown to be a spin-1 system with easy-plane anisotropy, 
exhibiting two SEs with opposite spin character in an external field, 
making it ideally suited to probe the tip polarization~\cite{ormaza2017a,verlhac2019a,supplmat}. 

Figure~\ref{fig:fig3}a shows the surface after the consecutive deposition of Fe islands, Nc molecules, and Co atoms. 
The ring-shaped Nc molecules are preferentially found 
at Cu vacancy edges which have formed during Fe deposition~\cite{brodde1993}. 
Fig.~\ref{fig:fig3}b shows the \ddidu\ signal recorded atop a Nc molecule at \unit{12}{T} with a spin-polarized tip. 
As sketched in the insets, the asymmetry of the spectrum is caused by the spin-polarized tunneling current and the transfer 
of angular momentum $\Delta S_z = \pm1$ between the tunneling electrons and the molecule in the SE process~\cite{loth2010}. 
Since the excitation increasing (decreasing) the molecule's spin is probed predominantly at negative (positive) bias voltage, 
we can conclude that this particular tip mainly probes the majority spin channel of the sample LDOS. 
The value of the spin polarization can be obtained by a fit to the experimental data displayed as a dashed line, 
yielding a polarization of $\approx$\,\unit{60}{\%} \cite{ternes2015a}. 

This 
pre-characterized tip is then used to perform spin-resolved 
measurements on Co atoms at 
\unit{12}{T}, where the splitting of the \ddidu\ features is maximal.
The resulting differential conductance and its derivative are displayed in Fig.~\ref{fig:fig3}c and d, respectively. 
For comparison, measurements with the same tip on bare Cu(111) are also displayed~\cite{supplmat}. 
Dashed blue and red lines indicate the position of the respective \ddidu\ peaks 
extracted from the spin-averaged measurements at the same field. 
It is evident, that the high-energy feature at positive bias (red) is prominently present, 
whereas the low-energy positive-bias feature (blue) is essentially absent. 
A clear statement concerning the dip at negative bias voltage (orange) is not possible, 
as the \ddidu\ signal atop the bare Cu(111) surface exhibits a dip of comparable intensity at a similar energy~\cite{supplmat}.
In combination with the analysis of the tip polarization on the Nc molecules, this finding allows us to ascribe 
the high-energy (red) peak a majority spin character, whereas we find that the low-energy (blue) peak has a minority spin character.

Against the background of theoretical and experimental work up to this date, this finding is surprising: 
If solely judged based on their energy shift in magnetic field, 
the various spectroscopic features in the Co atoms conductance signal would be attributed 
to states or excitations in the majority spin channel, as they shift to lower energies with increasing field strength (orange and blue peak), 
and in the minority spin channel, as they shift to higher energies with increasing field strength (red peak). 
Yet, at least for the positive-bias features (blue and red peak), the spin-polarized measurements 
reveal the opposite spin character of what is anticipated from their field-induced energy shift. 
This behavior again clearly contradicts the explanation of the ZBA's origin by the Kondo effect or common inelastic SEs. 

In contrast, the spinaron was explicitly predicted to shift to higher energies in an increasing magnetic field, 
while having a majority spin character, in agreement with the presented experimental data 
of the high-energy (red) feature in our STS measurements. 
Yet, the low-energy positive-bias feature, which similarly exhibits a field-induced shift 
contradicting its experimentally established spin character within conventional models, remains unexplained by the calculations 
published in Ref.~\cite{bouaziz2020}. Extending these calculations, 
which focused  on the $d_{z^2}$ orbital of the Co atom satisfying the symmetry conditions 
that enable tunneling of electrons to a tip positioned exactly atop the adatom, 
we here examine the electronic structure of all orbitals pertaining to the Co adatom in more detail.

\subsection*{Theory}
\vspace{-0.3cm}
\begin{figure*}[t]
\includegraphics[width=180mm]{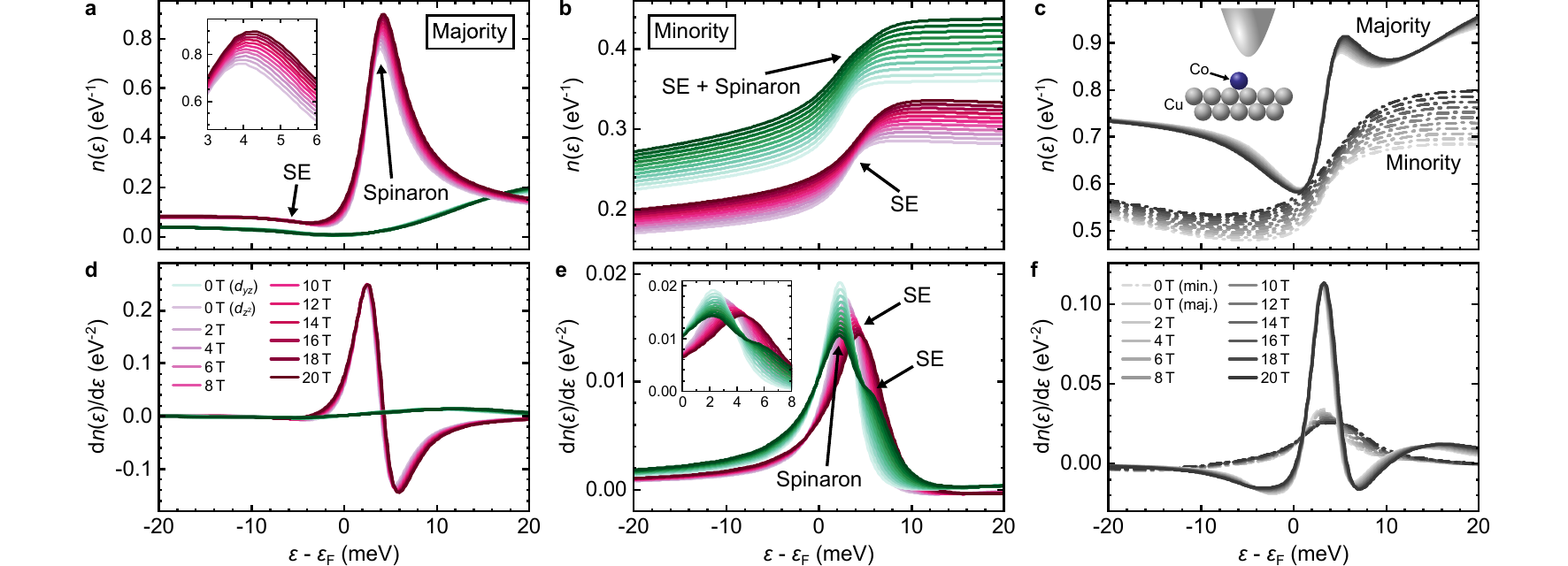}
\caption{\textbf{Orbital-resolved local density of states and theoretical inelastic tunneling spectra.} 
	\textbf{a} and \textbf{b}  Orbital-resolved local density of states (LDOS) at the adatom level of a Co atom on Cu(111) 
	for the $d_{z^2}$ and $d_{yz}$ (degenerate with $d_{xz}$) state 
	as function of the magnetic field for both the majority and minority spin channel. 	
	The presence of spinarons and intrinsic SEs can be identified by the characteristic field dependence of the visible features, 
	most clearly seen in the corresponding energy derivatives plotted in \textbf{d} and \textbf{e}. 
	The inset in \textbf{e} shows a zoom to the relevant features in the energy derivative of the minority spin channel LDOS. 
	The spin-resolved tunneling spectra and their energy derivatives are shown in \textbf{c} and \textbf{f}, 
	assuming a tip shifted slightly away from the Co atom's center.} \label{fig:fig4}
\end{figure*}

First-principles simulations indicate that the Co adatom 
carries a large spin (orbital) magnetic moment of $2\mu_{\rm B}$ ($0.47\mu_{\rm B}$), 
which points out-of-plane as favored by a significant magnetic anisotropy energy of \unit{4.3}{meV}. 
The latter induces gapped intrinsic SEs, which are expected to emerge 
in the adatom spectral function symmetrically around the Fermi energy~$\varepsilon_{\rm F}$. 
The interaction of the electronic states and the SEs renormalizes the electrons' self-energy as well as the electronic properties, 
which can be quantified via MBPT combined with relativistic TD-DFT (see Methods section). 

The resulting orbital-resolved local density of states (LDOS) of the Co adatom is presented in Fig.~\ref{fig:fig4}a,b  
in red (blue) for the $d_{z^2}$ ($d_{xz,yz}$) orbitals for different magnetic fields of up to 20 T. 
The respective energy derivatives $\dd n/\dd \varepsilon$ are plotted in Fig.~\ref{fig:fig4}d,e. 
From there, it is possible to deduce the characteristic response of each feature to the magnetic field, 
which allows to distinguish SEs and spinarons in our calculations as was discussed in the beginning of this paper.

In the majority spin channel, Fig.~\ref{fig:fig4}a,d, we identify the previously reported prominent $d_{z^2}$ spinaron 
at 4\,meV above $\varepsilon_{\rm F}$, which overlaps with the expected SE at $\varepsilon - \varepsilon_{\rm F} \approx - 4$\,meV.
In the minority spin channel, Fig.~\ref{fig:fig4}b,e, the $d_{z^2}$ orbital exclusively hosts a SE which is found at positive energy. 
The $d_{xz,yz}$ orbitals, equally exhibiting an intrinsic SE that shifts to higher energies with increasing field strength, additionally host a spinaron that shifts to lower energies~\cite{supplmat}. 
At zero field, both energetically coincide at $\approx 2.5$\,meV and visibly split in large magnetic fields due to their opposing response.

Based on the LDOS at the adatom level presented in Fig.~\ref{fig:fig4}a,b, 
we present the theoretical tunneling spectrum and its derivative 
obtained within the Tersoff-Haman approximation in Fig.~\ref{fig:fig4}c,f,
assuming a slightly shifted tip to capture the contribution of the $d_{xz,yz}$ orbitals~\cite{Tersoff1983}. 
In the experiment, tip orbitals deviating from the theoretically assumed $s$-wave tip could equally result 
in an enhancement of signatures from the $d_{xz,yz}$ orbitals in the tunneling spectrum. 
The resulting curves exhibit the known shape of the ZBA 
and will be compared to the experimentally obtained conductance in the following. 

\subsection{Discussion}
\vspace{-0.3cm}
Based on our \textit{ab-initio} calculations and their interpretation, we can now re-dissect our experimental results.
Staggeringly, our calculations support the presence of two spinaronic features in the LDOS of the Co atom on Cu(111), 
one being found in the majority spin channel of the $d_{z^2}$ orbital 
and the other one originating from the minority spin channel of the $d_{xz,yz}$ orbitals. 
Matching the experimentally determined spin character and direction of the magnetic field-induced energy shift 
as well as the energetic position at zero field of both features, 
we attribute the high-energy (red) feature in the experimental \ddidu\ signal to the $d_{z^2}$ spinaron in the majority channel 
and the low-energy (blue) feature to the $d_{xz,yz}$ spinaron in the minority channel.

Additional signatures of intrinsic SEs are predicted 
in all orbitals and spin channels as discussed above.
Experimentally, only the negative-bias (orange) \ddidu\ dip can presumably be related to a majority SE, 
although its spin character remains unresolved in our measurements. 
The two SEs predicted in the minority channel remain unresolved, 
potentially due to their energetic overlap with the intense spinaronic signals, 
which are also found in the calculated vacuum LDOS atop the Co atom and in the corresponding derivative in Fig.~\ref{fig:fig4}c,f.

Bearing in mind the \textit{ab-initio} nature of the theoretical framework that refrains from any fitting adjustment, 
the identified agreement between experimental and theoretical data is astounding, 
establishing the experimental findings as strong evidence for the presence of two spinarons in the LDOS of Co atoms on the Cu(111) surface.
Our work motivates further investigations to untwist the intriguing nature of the unraveled spinaronic states and 
to identify their presence in other sub-nanoscopic systems (e.g., other noble metal (111) surfaces, see Ref.~\cite{supplmat}). The presented study demands for the reassessment of previous work studying the interaction of the magnetic states in Co atoms with their environment~\cite{neel2010,bork2011} and inspires the exploration of the potential of engineering the spinarons' properties such as amplitude and spin-polarization via confinement effects or in atom-by-atom constructed nanostructures~\cite{chen1999,neel2008,neel2020,noei2023}. Finally, future research to establish the consequences of the presence of spinarons for spin-resolved transport could challenge the effects of magnetic impurities far beyond the field of surface physics.

\subsection{Methods}
\vspace{-0.3cm}
\textbf{Experiment.} The Cu(111) crystal is prepared by cycles of Ar-sputtering and annealing to \unit{500}{\degreecelsius}. 
Co atoms are deposited onto the surface from a Co rod at a substrate temperature of \unit{4.2}{K}. 
For spin-polarized measurements, first, bilayer high Fe islands are grown at room temperature with a surface coverage of \unit{25}{\%}. 
After cooling the sample to liquid helium temperature, Nc molecules are deposited 
by briefly placing the sample in front of a leak valve connected to the Nc container.
Finally, Co atoms are deposited \textit{in-situ}. 
Gently dipping the tip into the Fe islands leads to a transfer of Fe onto the tip apex, 
forming a magnetic cluster that results in a spin-polarization of the tunneling current.
All measurements are performed at \unit{1.4}{K}. 
The differential conductance data are recorded via a lock-in amplifier 
with a modulation frequency of $f = \unit{890}{Hz}$ and a modulation amplitude of $U_\mathrm{mod}^\mathrm{rms}=\unit{0.5}{mV}$.
Fitting of the spectra on Nc is performed within the dynamical scattering model introduced in Ref.~\cite{ternes2015a}.
\smallskip

\textbf{First-principles simulations.}
The ground state properties of the Co adatom deposited on Cu(111) 
were explored with density functional theory (DFT) as implemented 
in the scalar-relativistic full-electron Korringa-Kohn-Rostoker (KKR) Green function 
augmented self-consistently with spin-orbit interaction~\cite{Bauer:2014}. 
An angular momentum cutoff at $l_{\text{max}} = 3$ is assumed for the orbital expansion of the Green function 
and when extracting the local density of states a k-mesh of $300 \times 300$ is considered. 
The adatom relaxes towards the surface by $14\%$ of the Copper lattice parameter. 
The intrinsic SEs were obtained utilizing time-dependent DFT (TD-DFT)~\cite{Lounis:2010,Lounis:2011,Manuel:2015} 
including spin-orbit interaction, while the renormalized electronic states due to interaction with the SEs 
are evaluated within many-body perturbation theory (MBPT) accounting for relativistic effects~\cite{schweflinghaus2014,bouaziz2020}. 
Our results are found stable with both adiabatic local spin density and generalized gradient approximations.

The intrinsic spin excitation spectra are extracted from the dynamical magnetic susceptibility after solving the Dyson-like equation:
\begin{equation}
\underline{\chi}(\omega) = \underline{\chi}_\text{KS}(\omega) 
+ \underline{\chi}_\text{KS}(\omega)\,\underline{\mathcal{K}}\,\underline{\chi}(\omega)\quad,
\label{TDDFT_RPA}
\end{equation}
where the bare Kohn-Sham susceptibility $\underline{\chi}_\text{KS}(\omega)$ describes uncorrelated electron-hole excitations 
and $\underline{\mathcal{K}}$ corresponds to the exchange-correlation kernel. 
The electronic self-energy renormalized by the coupling of electronic states and SEs is extracted from MBPT, 
and consists of a convolution of the dynamical spin-susceptibility obtained from TD-DFT 
and Green functions calculated with DFT. More details are given in Refs.~\cite{schweflinghaus2014,bouaziz2020}.

\subsection{Acknowledgments}
\vspace{-0.3cm}
This research was supported by the DFG through SFB 1170 ``ToCoTronics'' and  
the W\"urzburg-Dresden Cluster of Excellence ct.qmat, EXC2147, project-id 390858490.
M.B., A.O.\ and F.F.\ thank Paolo Sessi (Max-Planck-Institut f{\"u}r Mikrostrukturphysik, Halle, Germany) 
for bringing this scientific topic to our attention.
S.L.\ acknowledges Sascha Brinker and Alexander Weismann for fruitful discussions.

\subsection{Author contributions}
F.F., A.O., M.B., and S.L. conceived the experiments. F.F. and A.O. conducted the measurements and analyzed the data. J.B. performed the first-principle simulations. S.L. and M.B. supervised the project. All authors
discussed the results. F.F. and S.L. wrote the manuscript with input from all authors.

\bibliography{References_v18}

\end{document}